\documentclass[aps,prd,twocolumn,showpacs,superscriptaddress,preprintnumbers, floatfix,nofootinbib]{revtex4-1}
\usepackage{amsmath}
\usepackage{hyperref, url}
\usepackage{graphicx,subfigure}
\usepackage[normalem]{ulem}
\usepackage{epsfig}
\usepackage{color,soul}
\usepackage{multirow}
\usepackage{array}
\usepackage[utf8]{inputenc}
\usepackage[T1]{fontenc}
\usepackage[dvipsnames]{xcolor}
\usepackage{comment}
\usepackage{aas_macros}
\hypersetup{colorlinks   = true,
            urlcolor     = blue,
            citecolor    = blue,
            linkcolor    = blue,
            menucolor    = blue,
            anchorcolor  = blue,
            filecolor    = blue}

\enlargethispage{\baselineskip}

\newcommand{\MScomment}[1]{}
\renewcommand{\MScomment}[1]{{\textcolor{red}{\bf [MS: #1]}}}

\usepackage{graphicx}

\newcommand{\nn}{\nonumber}

  \def\l{\lambda}  
 \def\r{\rho}

 \newcommand{\Ocal}{{\mathcal O}}

\begin{document}
\preprint{\hfill UCI-HEP-TR-2022-11}

\title{New Constraints on Dark Matter and Cosmic Neutrino Profiles through Gravity}

\author{Yu-Dai Tsai}
\email{yt444@cornell.edu}
\email{yudait1@uci.edu}
\affiliation{Department of Physics and Astronomy,
University of California, Irvine, CA 92697-4575, USA}
\affiliation{Fermi National Accelerator Laboratory (Fermilab), Batavia, IL 60510, USA}
\affiliation{Kavli Institute for Cosmological Physics (KICP), University of Chicago, Chicago, IL 60637, USA}

\author{Joshua Eby}
\email{joshaeby@gmail.com}
\affiliation{Kavli Institute for the Physics and Mathematics of the Universe (WPI), \mbox{The University of Tokyo Institutes for Advanced Study, The University of Tokyo, Kashiwa, Chiba 277-8583, Japan}}

\author{Jason Arakawa}
\email{arakawaj@udel.edu}
\affiliation{Department of Physics and Astronomy, University of Delaware, Newark, Delaware 19716, USA}
\affiliation{Department of Physics and Astronomy,
University of California, Irvine, CA 92697-4575, USA}

\author{Davide Farnocchia}
\email{Davide.Farnocchia@jpl.nasa.gov}
\affiliation{Jet Propulsion Laboratory, California Institute of Technology, Pasadena, CA 91109, USA}
\author{Marianna S. Safronova}
\email{msafrono@udel.edu}
\affiliation{Department of Physics and Astronomy, University of Delaware, Newark, Delaware 19716, USA}
\affiliation{Joint Quantum Institute, National Institute of Standards and Technology and the University of Maryland, College Park, Maryland 20742, USA}

\date{\today}

\begin{abstract}

We derive purely gravitational constraints on dark matter and cosmic neutrino profiles in the solar system using asteroid (101955) Bennu.
We focus on Bennu because of its extensive tracking data and high-fidelity trajectory modeling resulting from the OSIRIS-REx mission. We find that the local density of dark matter is bound by $\rho_{\rm DM}\lesssim 3.3\times 10^{-15}\;\rm kg/m^3 \simeq 6\times10^6\,\bar{\rho}_{\rm DM}$, in the vicinity of $\sim 1.1$ au (where $\bar{\rho}_{\rm DM}\simeq 0.3\;\rm GeV/cm^3$). We show that high-precision tracking data of solar system objects can constrain cosmic neutrino overdensities relative to the Standard Model prediction $\bar{n}_{\nu}$, at the level of $\eta\equiv n_\nu/\bar{n}_{\nu}\lesssim 1.7 \times 10^{11}(0.1 \;{\rm eV}/m_\nu)$ (Saturn), comparable to the existing bounds from KATRIN and other previous laboratory experiments (with $m_\nu$ the neutrino mass). These local bounds have interesting implications for existing and future direct-detection experiments. Our constraints apply to all dark matter candidates but are particularly meaningful for scenarios including solar halos, stellar basins, and axion miniclusters, which predict or allow overdensities in the solar system. Furthermore, introducing a DM-SM long-range fifth force with a strength $\tilde{\alpha}_D$ times stronger than gravity, Bennu can set a constraint on $\rho_{\rm DM}\lesssim \bar{\rho}_{\rm DM}\left(6 \times 10^6/\tilde{\alpha}_D\right)$. These constraints can be improved in the future as the accuracy of tracking data improves, observational arcs increase, and more missions visit asteroids.

\end{abstract}

\maketitle
\section{Introduction}
\label{sec:introduction}

The distribution of dark matter (DM) in the universe has been cemented as a crucial aspect of cosmology (see, e.g., \cite{SNLS:2005qlf,Planck:2018vyg}), as the ubiquitous gravitational influence of DM has driven much of the formation and dynamics of large structures such as galaxies and galaxy clusters. In our own galaxy, observations of stellar kinematics point towards an average density of  \(\overline{\rho}_{\text{DM}} \simeq 0.3\) GeV/cm\(^3\) near the position of the Sun \cite{Weber_2010,Nesti:2012zp,Bovy:2012tw,Read:2014qva}. However, there is no precise measurement of the density of dark matter in the solar system, and it may be much larger than the prediction from large-scale halo properties. Not only can substructure form in DM halos through gravity alone, overdensities of dark matter in the solar system are predicted or allowed by various beyond the Standard Model (BSM) theories (see, e.g., \cite{Vaquero:2018tib,Banerjee:2019epw,Anderson:2020rdk}). The interactions in these BSM theories that lead to overdensities are model-dependent, however, making it difficult to constrain the large swath of possibilities. It is therefore extremely important to develop model-independent, gravitational probes of dark matter \cite{Gron:1995rn,Anderson:1995dw,Khriplovich:2006sq,Sereno:2006mw,Frere:2007pi,Pitjev:2013sfa}, and to investigate dark matter models with only gravitational interactions (see, e.g., \cite{Kolb:2020fwh}). 

Additionally, detecting the cosmic neutrino background (C\(\nu\)B) is an important problem for neutrino physics and cosmology. The C\(\nu\)B decoupled from the early-universe plasma at temperatures of \(T\sim \Ocal(\text{MeV})\) \cite{Sellentin:2014gaa}, corresponding to when the universe was about a second old. Successful detection of the C\(\nu\)B, therefore, presents the ability to probe the very early universe. However, given its feeble interactions and non-relativistic energies, it is extremely difficult to directly detect the C$\nu$B, which has motivated attempts and proposals on the experimental front (see, e.g., \cite{Weinberg:1962zza}). At present, KATRIN \cite{KATRIN:2022kkv} has set the leading laboratory bound on the C$\nu$B density, and there are other ongoing experimental efforts such as PTOLEMY \cite{PTOLEMY:2018jst}. 
The constraints on the profiles of dark matter and cosmic neutrinos impact the physics interpretation of respective direct detection results, insofar as the density at the position of the experiment may be much larger than the value that is usually assumed (see, e.g., \cite{Banerjee:2019epw,Tsai:2021lly,Brdar:2022kpu}).

The possibility of probing the local dark matter density through purely gravitational interactions has been considered, mainly by exploring how an overdensity of dark matter in the solar system leads to an observable influence on the perihelion precession of the planets. In Ref.~\cite{Gron:1995rn,Anderson:1995dw}, weak limits were derived from the perihelion precession of Uranus, Neptune, and Pluto.
Refs.~\cite{Khriplovich:2006sq,Sereno:2006mw,Frere:2007pi} analyzed nearer planets to the Sun to set analogous constraints, which were improved in \cite{Pitjev:2013sfa} through the use of updated ephemerides; the most constraining of these were Mars and Saturn, which bounded the DM overdensity to be less than \(\Ocal(10^4)\) near the planets' respective orbits. These bounds could be extended to more objects and other orbital radii (including regions whose DM densities are so-far unconstrained by any direct measurement) by using asteroids. In fact, Ref.~\cite{Gron:1995rn} attempted to use the asteroid Icarus for this purpose, though the resulting limit was at the level of $10^8$ overdensity and neglected key systematic uncertainties (see Appendix \ref{app:uncert}).

Potentially hazardous near-Earth objects (NEOs) have motivated the careful tracking and modeling of asteroid orbits \cite{Atkinson00reportof}, for which observations and dedicated space missions (such as the OSIRIS-REx mission \cite{Lauretta_2017}) have recently provided exceptional orbital constraints \cite{FARNOCCHIA2021114594}.
In addition to planetary defense purposes, data from asteroids can be applied to study a wide range of fundamental physics topics including but not limited to general relativity (GR) \cite{1988A&A...200..248O, Verma:2013ata}, modified gravity \cite{Moffat:2006rq}, dark energy theory \cite{DeFelice:2009aj}, fifth forces \cite{Tsai:2021irw}, and the Yarkovsky effect \cite{Greenberg_2020}.
Planetary and asteroidal ephemerides are crucial not only for planetary defense and fundamental physics, but also for experimental applications, as they are also a source of uncertainty for the Pulsar Timing Array 
\cite{NANOGrav:2020tig}. 

When considering possible asteroids that could provide DM constraints, (101955) Bennu stands out. In fact, Bennu has a long ground-based data arc (from 1999 to 2020) and was the target of the OSIRIS-REx mission, which provided meter-level tracking data \cite{FARNOCCHIA2021114594}. Fitting these data required the most detailed force model ever implemented for an asteroid. To get a sense of the orbital constraints on asteroid Bennu, the Yarkovsky effect \cite{VOKROUHLICKY2000118} was measured with a signal-to-noise ratio of 1400, and the Poynting-Robertson \cite{Vok_Milani2000} drag was needed as part of the force model. Therefore, Bennu is an obvious candidate to look for constraints on DM using asteroids.

In this paper, we conduct the first analysis using asteroids to constrain the local DM profiles. Even in the absence of any additional interactions, the DM in the vicinity of the Sun causes the perihelion of objects orbiting the Sun to precess through purely gravitational interactions. By using orbital data and modeling, we derive an upper bound for the DM density \(\rho_{\rm{DM}}\) and cosmic neutrino density \(\rho_{\nu}\) near the orbits of asteroids by requiring that the DM-induced precession is within the variation due to measurement and modeling uncertainties.
We utilize the Comet and Asteroid Orbit Determination Package developed and maintained by NASA Jet Propulsion Lab (JPL), taking advantage of all the available observational data for Bennu to conduct our detailed analysis.

Limits on the matter density in our solar system can also be translated to a bound on the C\(\nu\)B. Although they are not yet sensitive enough to test the overdensities predicted by cosmic neutrino clustering models \cite{Stephenson:1996qj},
the planetary bounds are competitive with the KATRIN bounds, and constraints from asteroids can be improved in the future.

We also discuss models which introduce clustering mechanisms of DM and cosmic neutrinos, to motivate the types of specific models that could be constrained by our results. Even though our bound is model-independent in the sense that it only relies on gravity, the shape of the profile matters, and new shapes can be studied with our method. It is also worth noting that throughout this work, we assume that the cluster profile is spherically symmetric around the Sun; non-spherical distributions could lead to more dramatic effects.

The paper is organized as follows. 
We start in section~\ref{sec:clustering} by discussing models and mechanisms which induce dark matter or cosmic neutrino clustering.
In section~\ref{sec:DM}, we proceed to introduce the effect that invisible matter has on orbital precession and describe the detailed analysis we performed to derive the constraints.
Section~\ref{sec:results} describes the derived bounds on the local dark matter density; we also convert these results into bounds on the dark matter with a long-range fifth force and cosmic neutrinos.
We conclude in section~\ref{sec:summary}, and describe future directions to explore.
Unless otherwise specified, we use the convention of natural units ($\hbar = c = 1$) in this work.

\section{Clustering Mechanisms}
\label{sec:clustering}

\subsection{Dark Matter}
Under minimal cosmological assumptions, cold and non-interacting dark matter virializes into (approximately) spherically-symmetric halos. Under this assumption, observations of the circular velocity of stars in the Milky Way Galactic mid-plane imply a local DM density in the solar neighborhood of roughly $\bar{\rho}_{\rm DM} \simeq 0.3$ GeV/cm$^3$ (see, e.g., \cite{Weber_2010,Nesti:2012zp,Bovy:2012tw,Read:2014qva}).
Indeed, the vast majority of dark matter experiments use this value as an input to determine their sensitivity.

However, it is well-known that DM halos could possess substructure, which could arise from cosmological sources, or at late times from gravitational interactions alone. For non-interacting cold DM, the subhalo mass function follows an approximate power law, implying many subhalos over a wide range of physical scales in the galaxy (see e.g. \cite{Giocoli:2007uv}). Recent mergers of smaller halos into the Milky Way can also give rise to streams, a type of low-dispersion substructure, though the densities of streams are typically below the virial density in the halo (see, e.g., \cite{OHare:2019qxc}).

For dark matter composed of very light bosons (with $m_\phi \lesssim$ eV), the situation is even more complicated. Gravitationally-bound substructure is generally suppressed at scales below the de Broglie wavelength, $R\lesssim \lambda_{\rm dB} \sim 10^3{\rm \,km}\; (10^{-7}\,{\rm eV}/m_\phi)(10^{-3}/v)^2$, which can cut off the subhalo mass function at small masses \cite{Schutz:2020jox}. However, relaxation of the dark matter scalar fields can give rise to gravitationally-bound objects of size $R\simeq \l_{\rm dB}$, including self-gravitating boson stars \cite{Kaup:1968zz,Ruffini:1969qy,BREIT1984329,Colpi:1986ye,Seidel:1990jh,Friedberg:1986tq,Seidel:1991zh,Liddle:1993ha,Lee:1991ax,Chavanis:2011zi,Chavanis:2011zm,Eby:2016cnq} and bound bosonic halos \cite{Banerjee:2019epw,Banerjee:2019xuy,Tsai:2021lly} around large astrophysical objects.\footnote{Note also that gravitationally \emph{unbound} substructure, in the form of traveling waves sometimes known as \emph{quasiparticles} or \emph{granules} \cite{Hui:2016ltb,Bar-Or:2018pxz,Bar-Or:2020tys}, is ubiquitous in such models, though the overdensities in such waves are typically $\Ocal(1)$ of the background density.}
These high-density objects can modify the trajectories of planets and other objects in the solar system, thereby providing a possible method of detection.

In many theories of light DM scalars (including the QCD axion\footnote{See \cite{Weinberg:1977ma,Wilczek:1977pj,Dine:1981rt,Zhitnitsky:1980tq,Kim:1979if,Shifman:1979if}, or for more recent reviews see \cite{GrillidiCortona:2015jxo,DiLuzio:2020wdo} and references therein.}), the breaking of a high-scale $U(1)$ symmetry \cite{Peccei:1977hh} after inflation gives rise to large density perturbations that collapse at (or just before) matter-radiation equality \cite{Kolb:1993zz,Kolb:1993hw}, which are well-described by power-law profiles \cite{Ellis:2022grh}; similar arguments apply to spin-1 DM candidates \cite{Gorghetto:2022sue}. These overdensities are known as \emph{axion miniclusters}, and they may constitute a large fraction of the total DM mass in galaxies (see, e.g., \cite{Eggemeier:2019khm}). The precision of such predictions is limited due to the presence of tidal disruption, which can destroy these objects \cite{Dokuchaev:2017psd,Kavanagh:2020gcy,Xiao:2021nkb}, as well as uncertainties around the decay of topological defects (global strings and domain walls) after symmetry breaking \cite{Yamaguchi:1998gx,Gorghetto:2018myk,Vaquero:2018tib,Buschmann:2019icd,Gorghetto:2020qws}, which in part determines their abundance. However, 
it is possible that either a minicluster will pass through our solar system, modifying the local density for a finite (though potentially very long) time, or that the solar system formed in the presence of a minicluster.

Boson stars \cite{Kaup:1968zz,Ruffini:1969qy,Colpi:1986ye}, which can form in the cores of miniclusters \cite{Kolb:1993zz,Levkov:2018kau,Eggemeier:2019jsu,Kirkpatrick:2020fwd} or in more exotic scenarios \cite{Hertzberg:2020hsz,Kitajima:2021inh}, could modify the trajectory of asteroid orbits in a number of ways. First, a self-gravitating boson star could transit into our solar system, giving rise to a transient signal that might appear as a sudden kink in the trajectory of an asteroid. Secondly, boson stars with small masses could be bound inside the solar system, for example, inside the asteroid belt; this latter scenario would appear as an additional perturbing object in the analysis of asteroid trajectory. In this work, we do not investigate these possibilities, though they remain very interesting targets for future searches.

Dark matter bosons can also become bound in a bosonic halo around stars \cite{Banerjee:2019epw,Banerjee:2019xuy}, though the plausible range of densities for these objects remains the topic of ongoing investigations. 
Direct constraints on such bound overdensities around the Sun have been derived from observations of trajectories of planets, and this work provides analogous and competitive constraints using asteroids. Perihelion precession of planets gives rise to constraints roughly at the level of $(10^4 - 10^5)\bar{\rho}_{\rm DM}$ \cite{Pitjev:2013sfa}, assuming a bound solar halo with radius $R_\star \gtrsim 0.4$ au (the orbital radius of Mercury); in the bound halo scenario, this corresponds to a particle mass in the range $m_\phi \lesssim 10^{-14}$ eV \cite{Banerjee:2019epw,Banerjee:2019xuy}. Space-based missions 
with on-board quantum clocks flying nearer to the Sun may provide a highly-sensitive probe of smaller bound states in the future, and can reach DM scalar masses up to $m_\phi \simeq 10^{-13}$ eV \cite{Tsai:2021lly}.

Scalar and vector particles produced in the Sun which have speeds below the escape velocity can become captured, in what is called a \emph{solar basin} \cite{VanTilburg:2020jvl}. While these basins typically have densities below $\bar{\rho}_{\rm DM}$, in some scenarios, they can reach or even exceed this background value \cite{Lasenby:2020goo}. It has also been suggested that decay of axion quark nuggets can lead to captured axions near the Earth with densities at or near $\bar{\rho}_{\rm DM}$ \cite{Budker:2019zka}. Therefore local gravitational measurements can constrain these models as well.

One additional possibility of note is the formation of \emph{dark stars} from dark matter self-interaction \cite{Wu:2022wzw}. Although the formation typically involves early-Universe physics, certain parameters may result in late-time overdensity covering the solar system, thereby affecting the asteroid trajectories.

In each of the above models, overdensities beyond the galactic $\bar{\rho}_{\rm DM}$ are allowed by all existing constraints, which can greatly impact the direct detection of DM, both in existing and future experiments.

\subsection{Cosmic Neutrinos}

Within the SM, the \(\rm{C\nu B}\) has a predicted total number density of \(\bar{n}_\nu \simeq 336~\rm{cm}^{-3}\) today (counting all flavors of both neutrinos and anti-neutrinos). This, however, neglects the possibility of clustering. Since they are nonrelativistic, they are expected to cluster purely through gravitational influences. Moreover, if the neutrinos feel an attractive long-range BSM force, their clustering can be enhanced beyond the expectations from gravity alone. The overdensity is typically given in terms of the parameter \(\eta = n_{\nu}/\overline{n}_{\nu}\), where \(n_{\nu}\) is the number density of the cluster, and \(\overline{n}_{\nu}\) is the average density of relic neutrinos, here fixed to the SM prediction.

For instance, models which extend the SM to include a light scalar $\phi$ coupled to neutrinos via a Yukawa interaction \(y\phi \overline{\nu}\nu\) have been considered \cite{Smirnov:2022sfo}. Neutrinos are then subjected to a long-range attractive Yukawa potential and, depending on the parameters in the model, the neutrinos can cluster together as a Fermi gas. These clusters (sometimes referred to as \emph{neutrino stars}) have sizes that can be estimated by \(R
\sim 4\pi \sqrt{2}/(ym_{\nu})\), as long as the length scale of the interaction is much larger than the cluster size, \(m_{\phi} \ll 1/R\) (where $m_\nu$ is the neutrino mass). Neutrino clusters can therefore have length scales anywhere from \(\sim\rm{km}\) to \(10~\rm{Mpc}\). In this minimal extension, the overdensity can reach magnitudes of \(\eta = 10^7\).

Regardless of the clustering model, the overdensity of relic neutrinos can be constrained by direct observation. Before the KATRIN bounds, the best limits on \(\eta\) were given by LANL \cite{Robertson:1991vn} and Troitsk \cite{Lobashev:1999dv} neutrino mass experiments, and from cosmic ray data using the GZK limit \cite{Hwang_2005}. Currently, the KATRIN global upper limit of the neutrino overdensity at \(3\sigma\) is \(\eta = 7.7 \times 10^{11}\); the best fit limit ranges from \(2.6 \times 10^{11} \leq \eta \leq 3.8\times 10^{11}\) \cite{kellerer_2021}.

It should be noted that the Pauli exclusion principle would restrict C\(\nu\)B overdensities, assuming that they are free fermions, to the level of overdensity $\eta \sim 5\times 10^6 \, (m_{\nu}/ 1\, \rm{eV})^{3/2} \, (E_{F}/ 1.4 \,\mu\rm{eV})^{3/2} $, where \(E_F\) is the Fermi energy of the neutrino gas in the galaxy (see \cite{kellerer_2021}).
However, introducing new interactions for the neutrinos as described above can help alleviate this constraint.

\section{Gravitational Matter and Asteroid Trajectories}
\label{sec:DM}

\begin{figure*}
    \centering
    \includegraphics[width = .8\linewidth]{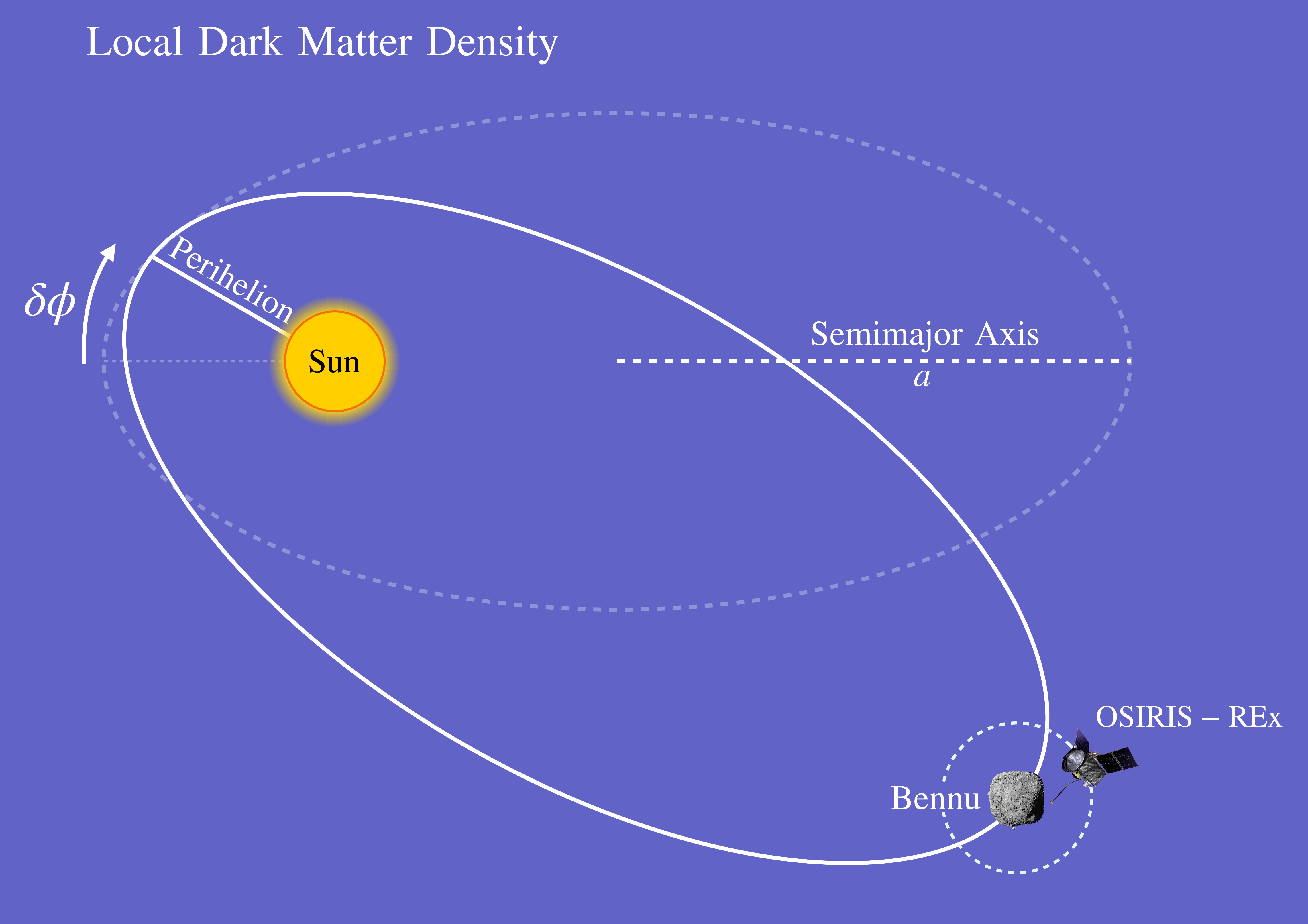}
    \caption{
    Schematic visualization of the perihelion precession of Bennu that would be caused by dark matter. Local gravitating matter, including dark matter or cosmic neutrinos, could cause presessions of asteroid orbits. Images of Bennu  \cite{bennu_2018} and OSIRIS-REx \cite{OSIRIS-REx_image} from NASA.}
    \label{fig:my_label}
\end{figure*}

Any matter\footnote{This section applies to dark matter, cosmic neutrinos, or any other gravitating matter around the Sun.} with gravitational effects in the solar neighborhood provides an additional non-\(1/r\) potential, which causes the orbits of objects in the solar system to precess. 
We derive below the orbital equations in the presence of the modified gravitational potential, and determine the magnitude of the resulting orbital precession, following the derivation given in Ref.~\cite{Gron:1995rn}. The assumption we make here is that the gravitating matter is non-relativistic and spherically symmetric around the Sun.
Note that in this section, we only consider gravitational effects, without any other interactions or fifth forces (see section \ref{sec:nongrav} for a model with non-gravitational interactions). 

For a static and spherically symmetric matter distribution, the gravitational metric takes the general form
\begin{align}
    ds^2 = -e^{2\mu(r)} dt^2 + e^{2\lambda(r)}dr^2 + r^2d\Omega^2,
\end{align}
\noindent where \(\mu(r)\) and \(\lambda(r)\) are functions determined by the Einstein field equations, \(G_{\mu \nu} = 8\pi G \,T_{\mu\nu}\), and \(d\Omega^2 = d\theta^2 + \sin^2{\theta}d\phi^2\) is the usual spherical line element. If we assume the density of the additional matter component \(\rho(r)\) is constant throughout the asteroid trajectory \(\rho(r) = \rho_0\), the solution to the Einstein equations is
\begin{align}
    ds^2 = &-\bigg(1 - \frac{2M_{\odot}}{r} + \frac{4\pi}{3}\rho_0 r^2\bigg)\, dt^2\nonumber\\
    &+ \frac{dr^2}{\bigg(1 - \frac{2M_{\odot}}{r} - \frac{8\pi}{3}\rho_0 r^2\bigg)} + r^2 d\Omega^2,
\end{align}

\noindent where \(M_{\odot}\) is the mass of the Sun. Given the above metric, the Lagrangian describing the motion of an object orbiting the Sun is
\begin{align}
    L = &-\frac{1}{2}\bigg(1 - \frac{2M_{\odot}}{r} + \frac{4\pi}{3}\rho_0 r^2\bigg)\, \dot{t}^2\nonumber\\
    &+ \frac{1}{2}\bigg(1 - \frac{2M_{\odot}}{r} - \frac{8\pi}{3}\rho_0 r^2\bigg)\, \dot{r}^2 + r^2 \dot{\phi}^2.
\end{align}
The spherical symmetry of the system allows us to restrict our equations to any plane (which we associate with the orbital plane of the probe mass), so we conveniently choose to restrict the motion to the \(\theta = \pi/2\) plane. Using the normalization condition for a test mass in this metric,
\begin{align}
    g_{\mu \nu} p^{\mu} p^{\nu} = -1,
\end{align}
with \(p_{\mu} = \partial{L}/\partial{\dot{x}^{\mu}}\), we have
\begin{align} 
    &p_{t} = \bigg(1 - \frac{2M_{\odot}}{r} + \frac{4\pi}{3}\rho_0 r^2\bigg)\, \dot{t}  \equiv E\nonumber\\
    &p_{r} = \bigg(1 - \frac{2M_{\odot}}{r} - \frac{8\pi}{3}\rho_0 r^2\bigg)\, \dot{r}\nonumber\\
    &p_{\theta} = r^2 \dot{\phi}  \equiv \ell,
\end{align}
where $E$ is identified with the energy of the orbit, and \(\ell\) the angular momentum.
Using the substitution \(u = 1/r\) yields the orbital equation,
\begin{align} \label{eq:orbital}
    \frac{d^2 u}{d\phi^2} + u = \frac{M_{\odot}}{\ell^2} + 3 M_{\odot} u^2 + \frac{4\pi}{3}\frac{\rho_0}{u^2 \ell^2}.
\end{align}

By perturbatively expanding around the Newtonian solution, \(u_0(\phi) = M_{\odot}(1 + e \cos{\phi})/\ell^2\), one can derive that (see Appendix~\ref{app:approx})
\begin{align}
    \delta \phi \simeq -4\pi^2 \rho_0 a^3 (1-e^2)^{1/2}/M_{\odot},
    \label{eq:approx}
\end{align}
\noindent which is the precession induced by a constant matter density along the asteroid trajectory, where $a$ is its semi-major axis. 
Note that if the profile of the gravitating matter near the asteroid trajectories is not constant, this would modify the metric, and the orbital equation would be modified accordingly. However, although the constant density assumption isn't entirely generic, it does naturally arise in many models, including a bound solar halo for very low DM masses \cite{Banerjee:2019xuy,Banerjee:2019epw,Tsai:2021lly}.

Note an important feature of the perihelion precession: $\delta \phi$ is proportional to the cube of the semi-major axis $a$, implying a benefit to the consideration of asteroids and Trans-Neptunian objects (TNOs) at larger distances from the Sun.

\subsection{DM Contributions to an asteroid's motion}

The motion of an asteroid is mostly driven by gravitational forces from the Sun, planets, Pluto, the Moon, and other perturbing asteroids \cite{FarnocchiaAst4}.
As observational arcs get extended and high-quality data collected, there are additional requirements on the accuracy of the force model and additional terms, such as solar radiation pressure \cite{Vok_Milani2000} and the Yarkovsky effect \cite{VOKROUHLICKY2000118}, that can be added.
Not only can these terms become necessary to match the observational data, but the parameters defining them can be estimated as part of the fit to the data \cite[e.g.,][]{Farnocchia2013}.

Bennu, the target of the OSIRIS-REx mission \cite{Lauretta_2017}, is the asteroid with the most-constrained trajectory in the entire catalog.
Besides ground-based optical data since 1999 and ground-based radar in 1999, 2005, and 2011, meter-level positional constraints were derived from OSIRIS-REx tracking data during the mission's proximity operations from January 2019 to October 2020 \cite{FARNOCCHIA2021114594}.
This exquisite dataset called for an unprecedented level of fidelity in the force model, including relativistic effects, perturbations from 343 asteroid perturbers, Yarkovsky effect based on a thermophysical model derived from in-situ characterization, solar radiation pressure, and Poynting-Robertson drag \cite{FARNOCCHIA2021114594}.
Therefore, Bennu stands out as the most promising asteroid candidate to derive constraints on dark matter.

Here, we use the same force model employed for the Bennu trajectory analysis based on OSIRIS-REx data \cite{FARNOCCHIA2021114594}.
We introduce an additional term to represent the perturbation due to dark matter.
In our derivation, as before, we only need to assume a constant and spherically-symmetric DM density populating radii from the perihelion, \(r_0\), to the aphelion, \(r_{\rm{max}}\), of the asteroid's orbit. For Bennu, \(r_0 = 0.90\) au and \(r_{\rm{max}} = 1.36\) au.
The mass contained within the orbit up to a radius $r$ is 
\begin{align}
    \mu (r)= 4\pi \int^r_{r_0}\rho_0 r^2 dr 
    = \frac{4\pi\rho_0}{3}(r^3-r_0^3).
\end{align}
The potential for an asteroid of mass \(m\) is 
\begin{align}
    U (r)= \frac{2\pi}{3} G m  \left[r_0^2 + 2 r_0^3 \left(\frac{1}{r_0} - \frac{1}{r}\right) - r^2\right] \rho_0,
\end{align}
which gives rise to a force of the form
\begin{align}\label{eq:force}
    \mathbf{F} (\mathbf{r})&= \frac{2\pi}{3} G m \rho_0 \left(\frac{2r_0^3}{r^2}-2r\right) \bf \hat{r}\nn\\
    &\simeq -\frac{4\pi}{3} G m \rho_0 r {\bf\hat{r}}
            + \frac{4\pi}{3} G m \rho_0 \frac{r_0^3}{r^2} {\bf\hat{r}}.
\end{align}
What matters here is the potential gradient in the radial direction during the asteroid orbit (the first term). Any mass contained inside \(r_0\) can be absorbed into the effective $1/r^2$ contribution of the Sun (the second term, neglected below); this additional mass is much smaller than that of the Sun. Therefore, the second term in Eq.~\eqref{eq:force} does not play an appreciable role and can be neglected.

\begin{figure*}
    \centering
    \includegraphics[scale=1.88]{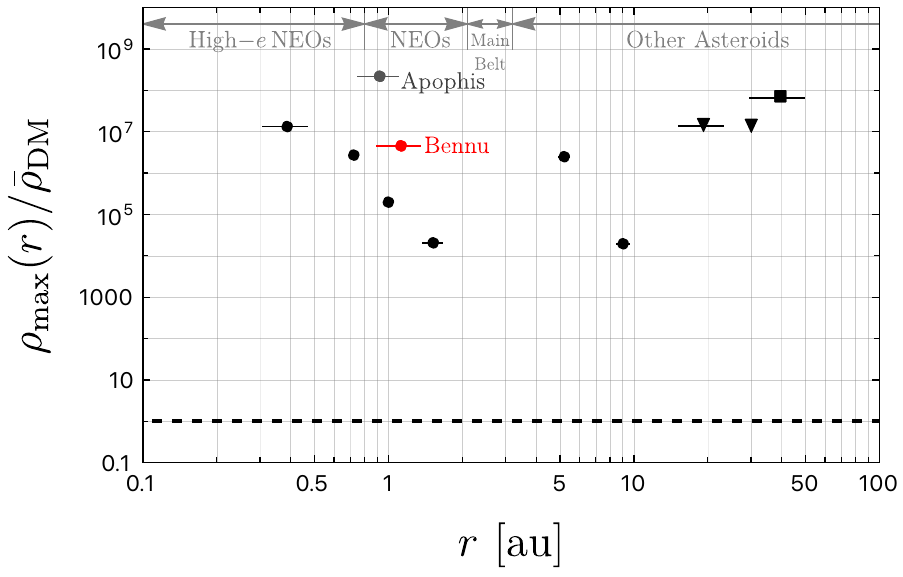}
    \caption{Asteroid (Bennu in red, Apophis and future targets in gray, as described in the text) and planetary (black square \cite{Gron:1995rn}, triangles \cite{Anderson:1995dw}, circles \cite{Pitjev:2013sfa}) constraints  on the overdensity of dark matter in the solar system. Each point probes the density of dark matter at that orbital distance; therefore they each provide distinct bounds; the length of the horizontal bars represents the range from $r_0$ to $r_{\rm max}$ probed by each object. Constraints on the profile of the dark matter can be determined by utilizing the full list of points together. The gray arrows represent the range of distances that could be probed by near-Earth objects (NEOs), including those with high-ellipticity, Main Belt objects in the range $2.1-3.4$ au, and other more distant asteroids, including trans-Neptunian objects (TNOs), as labeled.
    }
    \label{fig:densityconstraint}
\end{figure*}

\section{Results}
\label{sec:results}

\subsection{Purely Gravitational Constraints on Dark Matter}
We added the dark matter term in Eq.~\eqref{eq:force} to the force model and estimated $\rho_{\text{DM}}$ as part of the fit to the observational data.
For Bennu we obtain $\rho_{\text{DM}} = (-2.0 \pm 1.1) \times 10^{-15}$ kg/m$^3$ at \(1\sigma\), where the uncertainty given is the formal statistical uncertainty from the fit. 
The formal uncertainty in \(\rho_{\text{DM}}\) does not account for sources of error due to force model assumptions. We discuss and quantify the most important contributions in Appendix \ref{app:uncert}. Varying the model parameters yields 
differences up to \(\Delta \rho_{\text{DM}} \approx 2 \times 10^{-15}\) kg/m$^3$. Adding together the statistical uncertainty at \(3\sigma\) with the model 
sensitivity, the bound on the DM density is \(\rho_{\text{max}} \lesssim 3.3 \times 10^{-15}\) kg/m$^3$.
In Fig.~\ref{fig:densityconstraint}, we plot the constraint from Bennu (red) along with the previous constraints from planets (black) \cite{Gron:1995rn,Anderson:1995dw,Pitjev:2013sfa}, defining $\rho_{\rm max}$ as the maximum density $\rho_0$ allowed by the uncertainty of the measurement in both cases.
One can see that the constraints for Bennu are comparable to the level of the planets, and provide complementary results. 

Note that, again, our constraint and probes are model-independent in the sense that it does not rely on DM-SM interaction. However, the constraint would still be modified based on the assumption of the {\it shape} of the DM profile (more specifically, the radial dependence of the DM density).
One can easily modify the force in Eq. \eqref{eq:force} accordingly to produce a new bound for a non-trivial radial profile.

In the future, with sufficient precision, it could be possible to constrain the local dark matter density even in the absence of any local overdensity; in section \ref{sec:summary} we discuss some possibilities for improving these constraints. Note also that in the case of galactic dark matter, the variation in the profile on solar system scales is negligible, with a suppression factor of order \(\sim \rm{au}/\rm{kpc}\). This means that we can treat the dark matter profile near the Sun to be approximately constant, and the conditions laid out here still hold.

\subsection{Non-gravitational DM-SM Interactions}
\label{sec:nongrav}

Here, we discuss the constraints with the existence of DM-SM long-range interactions stronger than gravity.
Despite our results not relying on any non-gravitational DM-SM interactions, the constraints above do have important implications in the context of DM-SM long-range interactions stronger than gravity.
Such long-range force would enhance the effects on asteroids, and could also provide new mechanisms for the structure to form around the Sun.

We consider a Yukawa-type long-range force between DM and SM, mediated by a field $\phi$, with mass $m_\phi\ll \rm 1/au$. The mediator $\phi$ can be either a vector particle or a scalar particle, which will affect the sign of the dark fifth force. 
A simple parameterization can capture the complicated model-specifications in one parameter, $\tilde{\alpha}_D$, which is essentially a parameterization of how strong the DM-SM coupling is compared to gravity.
Under the same assumption that the DM forms a spherically-symmetric structure, the new long-range force is then parametrized as (c.f. Eq. \eqref{eq:force})
\begin{align}
    {\mathbf F}_{\rm DM-SM}({\mathbf r})&\simeq - \tilde{\alpha}_{D}\frac{4\pi}{3} G m \rho'_0 r \bf \hat{r}.
\end{align}

\noindent Requiring $F_{\rm DM-SM}(r)=F(r)$ from gravity in Eq.~\eqref{eq:force}, we arrive at a simple conclusion that
\begin{align}
\rho'_{\rm DM} (r)\lesssim \frac{\rho_{\rm max }(r)}{\tilde{\alpha}_D}.
\end{align}

\noindent Reading from Fig.~\ref{fig:densityconstraint}, for Bennu, this would imply that $\rho'_{\rm DM}\lesssim \left(6\times 10^6/\tilde{\alpha}_D\right)\bar{\rho}_{\rm DM}$ near the trajectory of Bennu, and $\rho'_{\rm DM}\lesssim \left(10^4/\tilde{\alpha}_D\right)\bar{\rho}_{\rm DM}$ near Saturn's trajectory.

We note that $\tilde{\alpha}_D$ is a heavy-handed parametrization, which does not describe the detailed DM model parameters. The class of models with such DM-SM interactions can be found in, e.g., \cite{Dutta:2019fxn, Baum:2022duc}, and will be further discussed in \cite{Tsai_2023}.
Also, note that the DM we consider here is quite different from the usual galactic DM, as galactic DM usually forms a structure centered around the Milky Way.
We simply show that with DM-SM long-range interactions, one can set stronger constraints on DM densities. For the usual galactic DM with long-range fifth forces to the SM, one can find more discussions in \cite{Sun:2019ico,Su:1994gu,Schlamminger:2007ht}.

\subsection{Constraints on Cosmic Neutrinos}
\label{sec:neutrinobound}

In the same light as probing the dark matter density in the solar system using the orbits of asteroids and planets, we can also constrain the overdensity of relic neutrinos in our solar system. If the average galactic density of dark matter is assumed to occupy the solar neighborhood, we can rescale the DM bounds into a bound on neutrino density.
The bound on a general overdensity is

\begin{equation}
    \eta \equiv \frac{n_\nu}{\bar{n}_\nu} \lesssim \frac{\rho_{\rm max}}{m_\nu \bar{n}_\nu} = 9\times10^6
    \left(\frac{\rho_{\rm max}}{\bar{\r}_{\rm DM}}\right)
    \left(\frac{\rm 0.1\;eV}{m_\nu}\right). \nn
\end{equation}

This results in bounds of 
\begin{align}
  \eta &\equiv \frac{n_\nu}{\bar{n}_\nu} \lesssim 1.7\times 10^{11}\left( \frac{\rm 0.1\;eV}{m_\nu}\right) \quad {\rm (Saturn)}, \nn \\
  \eta &\equiv \frac{n_\nu}{\bar{n}_\nu} \lesssim 5.4\times10^{13}\left( \frac{\rm 0.1\;eV}{m_\nu}\right) \quad {\rm (Bennu)}.
\end{align}
Recall from Section \ref{sec:clustering} that the current KATRIN global upper limit at \(3\sigma\) is \(\eta = 7.7 \times 10^{11}\). We can see above that the planetary bounds are competitive with such direct searches, and the asteroid bounds, while a bit weaker, could be improved in the future, as we discuss in Section~\ref{sec:summary}.

\section{Summary and Discussion}
\label{sec:summary}

In this work, we derived new constraints on the local density of dark matter and cosmic neutrinos using asteroids, via purely gravitational interactions. The physical picture is that a halo of invisible matter provides a perturbation that leads to orbital precession.
Using high-fidelity modeling of asteroid trajectories, we seek signals arising from gravitating matter.

We focused on Bennu, but our method can be applied to other asteroids,
allowing for new constraints on DM and cosmic neutrino profiles in the future. We determined bounds on the local density of DM in the vicinity of the orbit of Bennu (red point in Fig. \ref{fig:densityconstraint}):
\begin{align}
\rho_{\rm DM}(r\sim 1.1 \,\text{au})&\lesssim 3.3\times 10^{-15}\;\rm kg/m^3 \nn\\
        &\simeq 6\times10^6\,\bar{\rho}_{\rm DM}
        \; .
\end{align}
Although not at the same level of precision as Bennu, using the same method, we also derived constraints for another well-studied near-Earth asteroid, Apophis (see gray point in Fig. \ref{fig:densityconstraint}). This yields
\begin{align}
    \rho_{\rm DM}(r \sim 0.9\,\text{au})&\lesssim 1.53\times 10^{-13}\;\rm kg/m^3 \nn\\
    &\simeq 3\times10^8\,\bar{\rho}_{\rm DM} \; 
\end{align}
The constraint from Bennu is nearly two orders of magnitude better than that of Apophis, though the Apophis constraint may improve with data from OSIRIS-APEX \cite{2022LPICo2681.2011D}, the extended OSIRIS-REx mission.

The region marked by gray arrows in Fig. \ref{fig:densityconstraint} illustrates the possible reach for asteroid probes of the inner and outer solar system. In the inner region, high-ellipticity near-Earth objects (NEOs) can reach far inside the orbit of Mercury, whereas TNOs and other distant asteroids have radii extending beyond $10$ au; in between, thousands of Main Belt objects between $\sim 2-3$ au may be utilized. These probes may, however, be challenging to utilize. The inner region could see larger non-gravitational perturbations as well as additional sensitivity to contributions from relativistic effects and solar oblateness. Astrometric data of distant asteroids provide weaker positional constraints, and data arcs are much shorter relative to the orbital period, which also results in weaker orbital constraints.

Although we focused on Bennu, these results represent a proof of concept. There are many avenues to explore and ways to improve in the future. Using asteroids as a probe of local invisible matter has a number of benefits relative to planetary studies:
\begin{enumerate}
\item There are millions of objects to analyze, and although higher precision data is needed to provide stringent bounds, they could be useful in the future.
\item The trajectories of asteroids can span a much greater range of distances (especially when the eccentricity is large), both below $0.4$ au and above $9.5$ au (semi-major axis of Mercury and Saturn orbit, respectively). This allows one to constrain DM overdensities of vastly different sizes, and in regions 
where direct constraints are currently very weak or nonexistent (see also \cite{Tsai:2021lly}).
\end{enumerate}

Additionally, throughout the analysis, we assumed that the dark matter density is constant over the orbit of the asteroid. However, some models predict drastically different profiles that may change significantly over the orbits of asteroids with large eccentricities. This would modify the metric and, therefore, the orbital equations (as well as the force model), yielding a different form for the perihelion precession. Exploring these possibilities is left for future work.

The reliability of dark matter upper bounds can be improved by refining the force model using, e.g., ever-improved planetary ephemerides, better estimates of perturber masses, a more complete set of perturbing bodies, and modeling of other nongravitational effects such as solar wind.

Additionally, as the analytical estimate of the precession scales in Eq.~\eqref{eq:approx} as \(a^3\), asteroids that are further from the Sun could yield better bounds. For instance, the sensitivity reach using Jovian and Neptunian Trojans would be enhanced by a factor of \(\Ocal(10^2)\) and \(\Ocal(10^4)\) relative to Bennu, based on the relative distance to the Sun. If we obtain the same level of measurement and modeling precision for these Trojans as for Bennu, then a constraint would probe overdensities down to \(\Ocal(10^4)\) and \(\Ocal(10^2)\) respectively.

Future missions to other asteroids could prove helpful. Quantum sensors on a spacecraft, such as accelerometers, can improve navigation, which could lead to overall better asteroid tracking \cite{Rybak_1}. Additionally, detectors and quantum sensors stationed on asteroids could measure gravity gradients coming from dark matter or provide direct ranging, both of which have been considered in the context of gravitational wave (GW) detection \cite{Fedderke:2020yfy,Fedderke:2021kuy}.  Exploration of how space-based quantum sensors can improve asteroid science, in general, deems further investigation.

\section*{Acknowledgements}
We thank Vedran Brdar, Mariangela Lisanti, Marco Micheli, Sunny Vagnozzi, and Luca Visinelli for useful discussions. YDT is supported by U.S. National Science Foundation (NSF) Theoretical Physics Program, Grant PHY-1915005.
The work of JA was supported by NSF QLCI Award OMA - 2016244. The work of JE was supported by the World Premier International Research Center Initiative (WPI), MEXT, Japan, and by the JSPS KAKENHI Grant Numbers 21H05451 and 21K20366.
Part of this work was conducted at the Jet Propulsion Laboratory, California Institute of Technology, under a contract with the National Aeronautics and Space Administration (80NM0018D0004).
This work was initiated at the Aspen Center for Physics, which
is supported by the National Science Foundation grant PHY1607611.
YDT thanks the Kavli Institute for Theoretical Physics at the University of California, Santa Barbara, for its excellent program ``Neutrinos as a Portal to New Physics and Astrophysics,”
supported in part by the National Science Foundation
under Grant No. NSF PHY-1748958. YDT also thanks the Institute for Nuclear Theory at the University of Washington for its kind hospitality and stimulating research environment. This research was supported in part by the INT's U.S. Department of Energy grant No. DE-FG02- 00ER41132.
YDT dedicate this work to Ping-Kun Tsai.

\appendix

\section{Uncertainties in Asteroid Modeling}
\label{app:uncert}

As discussed by Ref~\cite{FARNOCCHIA2021114594}, there are force model limitations, and we should ensure that the results are not affected by modeling assumptions. 
Within the high-fidelity model, we vary the assumptions individually, and rerun the fit to the dark matter density. In Table \ref{table:uncert}, we show the extracted values of the 
\(\rho_{\text{DM}}\) compared to the baseline model. The assumption that yields the largest change is using a more recent ephemeris (DE440 instead of the DE424 baseline), which we report in our constraint.  Otherwise, the results are not too sensitive to the modeling assumptions. There are some effects that are not captured in the model, such as solar winds, YORP effect, ejected particles, and interactions between Bennu and the OSIRIS-REx spacecraft. Of these, solar winds are likely to be the most important. As pointed out in Ref.~\cite{Pitjev:2013sfa}, solar winds are approximately spherical with a density of roughly \(\Ocal(10^{-20})\) kg/m\(^3\) near the Earth's orbit. Since the contribution that leads to precession depends on the gradient of the potential, the much smaller density at these orbits renders the solar-wind effects negligible. 

Lastly, we checked that the 343 perturbers that we use in the model lead to a sufficiently convergent result. As a check, we determine \(\rho_{\rm DM}\) while adjusting the number of perturbers included in the analysis, and find that the density prediction does indeed converge well, as displayed in Table \ref{table:perturbers}.

\begin{table}[t]
\centering
\setlength{\tabcolsep}{6pt}
\begin{tabular}{c  c} 
 \hline
 Model Parameter & \(\Delta \rho_{\text{DM}}\) [kg/m\(^3\)] \\
 \hline\hline
 \vspace{.1cm}
 DE440 & \(2.0\times 10^{-15}\)\\
 DE440 w/ TNOs & \(1.8\times 10^{-15}\) \\
 Galilean satellite separation & \( -9.9\times 10^{-17}\)\\
 Solar quadruple moment $J_2$ & \(2.4\times 10^{-16}\) \\
 No PR drag & \(-1.0\times 10^{-23}\)\\
 Solar mass loss & \(-8.4\times 10^{-20}\)\\
 Linear Yarkovsky model (V00) & \( 1.5\times 10^{-15}\)\\
 Non-spherical SRP & \(-1.0\times 10^{-17}\)\\
 \hline
\end{tabular}
\caption{List of the modeling assumptions that were adjusted, and the resulting change in the extracted values of the dark matter density. The baseline is given by  \(\rho_{\rm{DM}} = (-2.0 \pm 1.1)\times 10^{-15}\) kg/m\(^3\).
DE440 stands for the replacement of the ephemeris DE424 with DE440. DE440 w/ TNOs means the additional inclusion of the TNOs in the force model. In the table, we also consider the effects of the Galilean satellites being treated as separate objects from Jupiter, adding the solar
quadruple moment $J_2$, turning off the Poynting–Robertson (PR) drag, adding solar mass loss, switching to the linear Yarkovsky model, and adding non-spherical Solar Radiation Pressure (SRP).
}
\label{table:uncert}

\end{table}

\begin{table}[t]
\centering
\setlength{\tabcolsep}{6pt}
\begin{tabular}{c  c} 
 \hline
 Number of Perturbers & \(\rho_{\text{DM}}\) [kg/m\(^3\)] \\
 \hline\hline
 \vspace{.1cm}
 16 &  \(-7.27\times 10^{-16}   \pm 6.76\times 10^{-16}\) \\
 32 & \(-1.00\times 10^{-15}  \pm  7.49\times 10^{-16}\)\\
 64 & \(-1.79\times 10^{-15}   \pm 9.78\times 10^{-16}\) \\
 128 & \( -1.92\times 10^{-15} \pm  1.03\times 10^{-15}\)\\
 256 & \(-2.02\times 10^{-15}   \pm 1.06\times 10^{-15}\) \\
 343 & \(-2.04\times 10^{-15}   \pm 1.06\times 10^{-15}\)\\
 \hline
\end{tabular}
\caption{Dark matter density determined by the fit as a function of the number of perturbers included in the model. As we approach 343 perturbers, \(\rho_{\text{DM}}\) and its uncertainty converge, implying that the number of perturbers included is sufficient.}
\label{table:perturbers}
\end{table}

\section{Analytical Approximation for Precession from Dark Matter}
\label{app:approx}

Here we rederive the expression for the dark matter induced perihelion precession, starting from the orbital equations of motion.

First, recall the orbital equation of Eq. \eqref{eq:orbital}:

\begin{align}
\label{eq:num}
    \frac{d^2 u}{d\phi^2} + u = \frac{M_\odot}{\ell^2} + 3M_\odot u^2 + \frac{4\pi}{3} \frac{\rho_0}{u^3 \ell^2}.
\end{align}

\noindent Perturbatively expanding around the Newtonian solution, \(u_0(\phi) = M_\odot(1 + e \cos{\phi})/\ell^2\) (where $e$ is the eccentricity), and writing our solution as \(u(r) = u_0(r) + \Delta u(r)\), gives

\begin{align}
    \frac{d^2 \Delta u}{d\phi^2} + \Delta u &=  3M_{\odot} u_0^2 + \frac{4\pi}{3} \frac{\rho_0}{u_0^3 \ell^2}\\
    &= \frac{3M_{\odot}}{\ell^4}(1+e\cos{\phi})^2\nn\\ 
    &\,\,\,+ \frac{4\pi}{3}\rho_0 \frac{\ell^4}{M_{\odot}^3} \frac{1}{(1+e\cos{\phi})^3}.
\end{align}
The solution, keeping only terms that contribute to precession, is
\begin{align}
    \Delta u &= \frac{3M_{\odot}^3}{\ell^4} e \phi \sin{\phi}\\ 
    &- \frac{4\pi \rho_0 \ell^4}{M_{\odot}^3} \frac{e}{(1-e^2)^{5/2}} \arctan{\bigg(\frac{1-e}{\sqrt{1-e^2}} \tan{\frac{\phi}{2}}\bigg)} \sin{\phi}. \nn 
\end{align}
If we expand the \(\arctan\) term around \(e=0\), only the zeroth-order term accumulates over the period of the orbit. Keeping only this term yields 
\begin{align}
    \Delta u &= \frac{3M_{\odot}^3}{\ell^4} e \phi \sin{\phi}- \frac{2\pi \rho_0 \ell^4}{M_{\odot}^3} \frac{1}{(1-e^2)^{5/2}} e\phi \sin{\phi}.
    \label{eq:deltau}
\end{align}
Note that these perturbations vanish when \(e \rightarrow 0\), 
which implies that the radial distance 
must change throughout the orbit to see precession. 

We can recast this result into a change in the \(\phi\)-period \(P\) of the orbit, since, as the perihelion precesses, the object will return to its perihelion at an angle that is different than \(2\pi\). That is, we write the full solution as
\begin{align}
    u(\phi) \approx \frac{M_{\odot}}{\ell^2} \bigg(1+e\cos{(\phi \{1-\alpha\})}\bigg),
\end{align}

\noindent where \(\alpha\) shifts the \(\phi\)-period. For our perturbative expansion \(\alpha \ll 1\), so we expand this as \(\cos{(\phi \{1-\alpha\})} \approx \cos{\phi} + \alpha \phi \sin{\phi}+\Ocal(\alpha^2)\). We therefore identify the coefficients on \(\phi \sin{\phi}\) of Eq.~\eqref{eq:deltau} with \(\alpha M_{\odot}/\ell^2\), giving
\begin{align}
    \alpha = \frac{3M_{\odot}}{\ell^2} - \frac{2\pi \rho_0 \ell^6}{M_{\odot}^4} \frac{1}{(1-e^2)^{5/2}}.
\end{align}
\noindent In terms of \(\alpha\), the \(\phi\)-period becomes
\begin{align}
    P = \frac{2\pi}{1-\alpha} \approx 2\pi(1+\alpha).
\end{align}
The precession \(\delta \phi\) is defined by how much the $\phi$-period is shifted from \(2\pi\), i.e. \(\delta \phi = P - 2\pi\), which implies
\begin{align}
\label{eq:app}
    \delta \phi &= 2\pi \alpha\nn\\
    &= \frac{6\pi M_{\odot}}{a(1-e^2)} - \frac{4\pi^2 \rho_0}{M_{\odot}} a^3 \sqrt{1-e^2},
\end{align}
where we have used \(a = M_{\odot}/(\ell^2 (1-e^2))\). The second term reproduces Eq.~\eqref{eq:approx}, as expected. We have checked this result numerically by solving Eq.~\eqref{eq:num} directly, and we recover Eq.~\eqref{eq:app} for small but finite $e$.

\bibliographystyle{JHEP}
\bibliography{ref}

\end{document}